\documentclass[twocolumn,10pt,amsmath,amssymb,prc,aps,superscriptaddress, nofootinbib, longbibliography]{revtex4-1}

\usepackage[T1]{fontenc}
\usepackage[utf8]{inputenc}
\usepackage{xargs}
\usepackage{graphicx}
\usepackage{bm}
\usepackage{hyperref}
\hypersetup{
 colorlinks  = true, 
 urlcolor   = black, 
 linkcolor  = blue, 
 citecolor  = red  
} 

\newcommand{\WUT}{\mbox{Faculty of Physics, Warsaw University of Technology, Ulica Koszykowa 75, 00-662 Warsaw, Poland}}
\newcommand{\UW}{\mbox{Department of Physics, University of Washington, Seattle, WA 981951560, USA}}
\newcommand{\IFPAN}{\mbox{Institute of Physics, Polish Academy of Sciences, Aleja Lotnikow 32/46, PL-02668 Warsaw, Poland}}

\newcommand{\Ebsk}{E_\textrm{BSk}}
\newcommand{\Epsilonbsk}{\mathcal{E}_\textrm{BSk}}
\newcommand{\cm}{\textrm{c.m.}}
\newcommand{\Zcm}{Z_\textrm{c.m.}}
\newcommand{\Mcm}{M_\textrm{c.m.}}
\newcommand{\Rcm}{\bm{R}_\textrm{c.m.}}

\newcommand{\Up}{U_\textrm{p}}
\newcommand{\UpBsk}{U_{\textrm{p},\textrm{BSk}}}
\newcommand{\gammaCM}{\gamma_\cm}

\newcommand{\Vlandau}{v_\textrm{l}}

\newcommand{\MeV}{\textrm{MeV}}
\newcommand{\fm}{\textrm{~fm}}
\newcommand{\kF}{k_\mathrm{F}}

\begin{document}

\title{Dynamical scheme for computing the mass parameter of a system in a medium}

\author{Agata Zdanowicz} 
\affiliation{\WUT}

\author{Daniel P{\k e}cak} \email{daniel.pecak@ifpan.edu.pl}
\affiliation{\WUT} \affiliation{\IFPAN} 

\author{Piotr Magierski} \email{piotr.magierski@pw.edu.pl} 
\affiliation{\WUT} \affiliation{\UW}  

\author{Gabriel Wlaz\l{}owski} \email{gabriel.wlazlowski@pw.edu.pl} \affiliation{\WUT} \affiliation{\UW} 

\date{\today}
\begin{abstract}
We present a new method for extracting a mass parameter using time-dependent density functional theory for an arbitrary physical system, provided the adiabatic limit is achievable. This approach works for collective variables also in the presence of a medium, in particular for the nuclei interacting with a neutron background. We apply the method to extract mass parameters of impurities in the neutron star crust, like their inertial masses and quadrupole mass parameters. The extracted mass parameters at various depths of the inner crust are compared with other methods, including the hydrodynamic approach. The presented method opens avenues for the construction of an effective model of neutron star crust grounded in microscopic calculations.  
\end{abstract}
\maketitle

\section{Introduction}
Effective models are indispensable tools in physics that inevitably emerge for complex systems. With their help, one can connect a physical phenomenon taking place across various scales. Whenever we change degrees of freedom and construct a new effective model operating at larger length scales, input from the underlying model is required. In this way, we obtain a hierarchy of models spanning from macroscopic approaches through mesoscopic ones up to microscopic ones. As an example, relevant to this paper, let us consider the problem of constructing an effective model of a rotating neutron star. At the largest scales, it can be modeled by means of hydrodynamical approaches~\cite{Graber2017,PhysRevD.95.083005}. To explain observed phenomena of neutron star glitches (sudden decreases of rotation period), one needs to revert to a multifluid picture, with the extra assumption that one of the fluids is in a superfluid state~\cite{Haskell2015,Khomenko2018}. Besides the equation of state, such models require various macroscopic parameters to be provided, like viscosities, conductivities, and mutual friction parameters between the fluid components. Going down in the hierarchy, by increasing the resolution, we enter the regime of mesoscopic models, where we start to resolve quantized vortices present in the superfluid component, which coexists with the so-called {\it impurities} created by the other fluid component~\cite{Howitt2020,Cheunchitra2024}. Precisely, neutron matter forms the superfluid component, filled out with vortices, while protons form various structures, which for the majority of the inner crust have spherical shapes. Electrons, which are essential for charge neutrality, are well approximated by their uniform distribution. At this level, one usually refers to the Vortex Filament Model (VFM), which was originally developed for modeling vortices in superfluid helium~\cite{Tsubota2013}. It is a quasiclassical model, where one writes Newton's equations for quantum vortices (presented as filaments in 3D or points in 2D) that interact with all other elements of the system. The drag force exerted on vortices by the impurities is the origin of the mutual force that appears in the hydrodynamical model. Increasing the resolution further, we start to resolve the building blocks of the neutron star matter: neutrons, protons, and electrons (and possibly other elementary particles deep in the star's core). At this level, one needs to revert to the microscopic description grounded in many-body quantum mechanics. The quantized vortices turn out to be a collective flow of neutrons around some line in space, while impurities correspond to regions where protons (together with neutrons) concentrate into various structures~\cite{Negele1973,pearson2018}, usually into nuclei-like objects, but more complex structures are also allowed~\cite{RevModPhys.89.041002}. The effective interaction between the flow of neutrons and the bound state of protons is the origin of the drag force that enters the mesoscopic models. 

The process of extracting effective model parameters from the underlying approach is a complex problem. Here, we focus on the method of extracting one of the most fundamental quantities that enters into effective models, namely \textit{mass parameters}. Basically, each effective model contains a term that characterizes the inertia of the selected degree of freedom $Q$. Schematically, the effective Hamiltonian can be written  
\begin{equation}
H_{Q} \approx \frac{M_Q\dot{Q}^2}{2} + V(Q),
\label{eq:HQ}
\end{equation}
where $Q$ is the selected collective variable.
Such a form is achievable once the adiabatic limit can be recovered in the system, which means that the terms describing couplings to other degrees of freedom can be neglected~\cite{Baranger1978,villars,goeke}. 
The first term is the kinetic term that is quantified by the mass parameter $M_Q$, and $V$ stands for the potential energy. The choice of the collective variable is motivated 
by physical intuition, although a more precise definition is also possible. In the latter case, one searches for the collective path (or space),  which is maximally decoupled from 
the other degrees of freedom (see~\cite{PhysRevC.109.L051301, PhysRevC.105.034603, RevModPhys.88.045004} and references therein).
This can be important in the case where the choice of collective variables is not obvious, as in the case of exotically deformed nuclei forming the pasta phase. Whatever method is applied to identify $Q$, it implicitly assumes that the adiabatic limit can be achieved, in which the motion of the system is governed by Hamiltonian~(\ref{eq:HQ}).

\begin{figure}[b]
 \includegraphics[width=0.95\columnwidth]{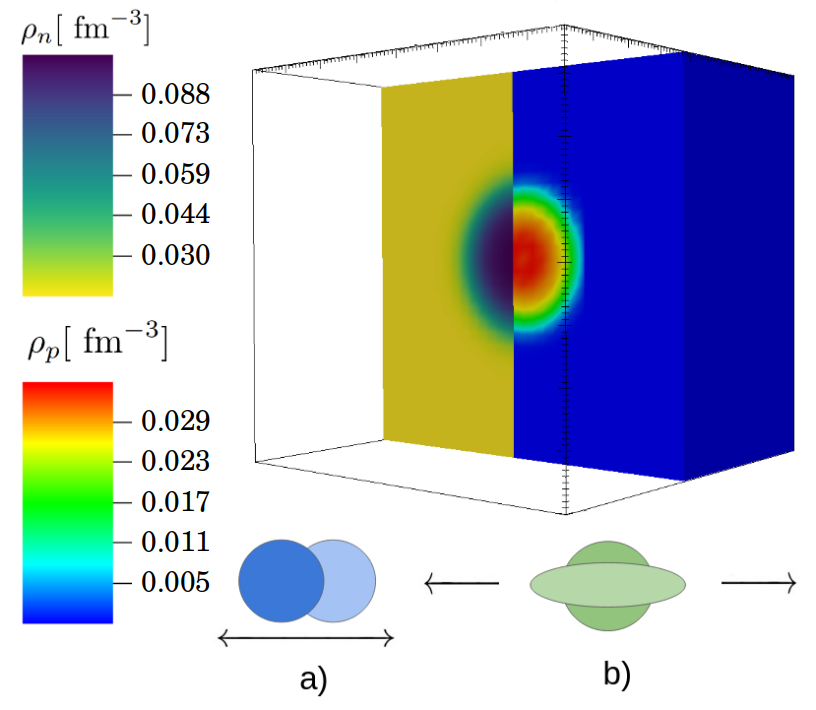}
 \caption{Scenario considered in this work: a nucleus ${}_{40}$Zr immersed in a superfluid neutron matter. The simplest types of collective modes that nuclei can execute are: (a) center of mass and (b) quadrupole moment oscillations. From the time evolution of  the collective degrees of freedom associated with the mode, we can fit the parameters of interest, such as the mass parameter and dissipation coefficients.
 }
 \label{fig:sketch}
\end{figure}

In this paper, we present a general-purpose method for extracting mass parameters for nuclear systems by means of the time-dependent density functional theory approach (TDDFT), which is considered as an efficient tool to investigate various problems of nuclear dynamics at low energies at the microscopic level \cite{RevModPhys.88.045004}. In the field of neutron stars, TDDFT has been applied e.g. in examination of nuclear pasta phase  \cite{schuetrumpf}, entrainment effects in the inner crust \cite{sekizawa}, pairing correlations \cite{stetcu} or development of band theory \cite{kashiwaba-band-theory}. While there are methods for extracting the effective mass parameters, as shown e.g. in \cite{umar,simenel}, most of them rely on various assumptions, and they are typically designed to be applicable for extracting the mass parameters in a vacuum (see, e.g., the review paper~\cite{Li2022}).
The presented approach is free of these limitations, and we will demonstrate it for the nuclear impurity immersed in superfluid neutron liquid, i.e., for the conditions relevant to the neutron star crust. 
The calculations will be delivered for mass parameters associated with translational motion and quadrupole vibrations. The selected collective variables are schematically presented in Fig.~\ref{fig:sketch}. 

The article is structured as follows: in Sec.~\ref{sec:method}, we introduce the aim and context of this work. In Sec.~\ref{sec:results}, we apply the method to two collective variables: the center of mass and the quadrupole moment. Finally, in Sec.~\ref{sec:conclusions}, we conclude the paper.

\section{Method}
\label{sec:method}

To make the problem more concrete, let us consider a Wigner crystal of neutron-rich nuclei in the crust of a neutron star. These nuclei can oscillate within the crystal, as illustrated in Fig.~\ref{fig:sketch}(a). The collective variable $Q$ can then be associated with the impurity’s position $\Rcm$, which, in a microscopic description, corresponds to the center of mass of the protons. The potential energy in this system is primarily due to Coulomb interactions between impurities. Even in this relatively simple setting, determining the effective mass of an impurity immersed in neutron matter remains a challenging task. A variety of approaches have been developed to address this problem~\cite{magierski2004medium, magierskibulgac2004, martin2016superfluid, chamel2013,pecak2024WBSkMeff}. The Wigner crystal is also expected to be threaded by quantum vortices. Work~\cite{wlazlowski2016vortex} demonstrated that the interaction between an impurity and a vortex can induce a quadrupole deformation of the impurity, Fig.~\ref{fig:sketch}(b). This suggests that the quadrupole moment $Q_{20}$ could also serve as a relevant collective variable in the effective model. It is also possible that higher-order nuclear multipole moments play a significant role.

The shape parametrization given by
\begin{equation}\label{eq:Rthetapshi}
R(\theta,\phi) = R_0 \left( 1+\sum_{\lambda=0}^{\infty}\sum_{\mu=-\lambda}^{\lambda} \alpha_{\lambda\mu} Y_{\lambda\mu}(\theta,\phi) \right),
\end{equation}
is commonly used in effective (collective) models to describe surface deformations of an impurity. Here, $R_0$ denotes the radius of the undeformed configuration, and the coefficients $\alpha_{\lambda\mu}$ represent collective degrees of freedom associated with shape oscillations. In this framework, each $\alpha_{\lambda\mu}$ corresponds to a collective mode with its own mass parameter. For example, $\lambda=1$ describes dipole (center-of-mass) motion, while $\lambda=2$ corresponds to quadrupole deformation.
While this form suggests an explicit impurity with a sharp boundary, we stress that the aim is to extract the related mass parameters from a microscopic approach, where the nucleus and the surrounding neutron medium are treated self-consistently on equal footing. The parametrization is introduced here solely to define and motivate the collective coordinates used in the analysis, and to provide a connection to the effective models where the concept of a shape is well defined.

In order to investigate the dynamics of the many-body system, we will consequently use an approach based on the density functional theory (DFT). In the case of nuclear systems, it offers a computationally tractable method, where the energy functional is constructed using various types of densities relevant to the problem. In recent decades, it has become one of the standard theoretical tools to study medium and heavy nuclei~\cite{RevModPhys.88.045004,Col2020,Yang2020}.
The heart of each DFT method is an energy functional
\begin{equation}
    E(t) = E[\rho(\bm{r},t), \ldots],
    \label{eq:Et}
\end{equation}
which maps the density of particles $\rho$ at a given position $\bm{r}$ and time $t$ into instantaneous energy (generally, the functional can also depend on other densities marked by dots). We assume that the density is also sufficient to define a collective variable
\begin{equation}
    Q(t) = Q[\rho(\bm{r},t)].
    \label{eq:QT}
\end{equation}
The aim is to construct the effective Hamiltonian for the collective variable, which means that we would like to approximate the energy~(\ref{eq:Et}) by a form given by Eq.~(\ref{eq:HQ}). Suppose we have at hand a microscopic method that can provide time evolution of the density $\rho(\bm{r},t)$, and related  quantities, like $Q(t)$.
If we restrict our considerations to cases where the collective variable exhibits small fluctuations $\delta Q(t) = Q(t)-Q_0$ with respect to its equilibrium value $Q_0$, then the effective potential can be well approximated as
\begin{equation}
    V(Q)\approx \frac{1}{2}M_Q\omega_Q^2 \left(Q(t)-Q_0\right)^2.
\end{equation}
In such a case, the recorded dynamics of $Q(t)$ should follow harmonic oscillations with frequency $\omega_Q$. The measured frequency provides us with information about the mode's energy $ E_Q = \hbar \omega_Q$. However, it is insufficient to deliver information about the mass parameter $M_Q$. But we can consider dynamics arising from a modified functional, supplemented by an additional harmonic potential that couples to the collective variable 
\begin{equation}
    E^\prime(t) = E(t) + \frac{1}{2}k \left( Q(t)-Q_0\right)^2.
    \label{eq:Eprimet}
\end{equation}
Upon this modification, we can rearrange the terms so that the effective potential takes the form
\begin{equation}
    V^\prime(Q)\approx \frac{1}{2}M_Q \omega^2(k) \left(Q(t)-Q_0\right)^2,
\end{equation}
where
\begin{equation}
    \omega^2(k) = \omega_Q^2+\frac{k}{M_Q},
    \label{eq:omegak}
\end{equation}
and the collective variable will oscillate with modified frequency $\omega(k)$ that is sensitive to the mass parameter $M_Q$. The procedure for extracting the mass parameter is then as follows:
\begin{enumerate}
    \item Solve time-dependent equations of motion arising from the modified energy functional $E^\prime$, Eq.~(\ref{eq:Eprimet}), for different values of parameter $k$ and for the case corresponding to the small amplitude oscillations in the variable $Q(t)$.\\
    \item Extract frequency of oscillations $\omega(k)$ and fit to it formula~(\ref{eq:omegak}) with respect of parameters $\omega_Q$ and $M_Q$. 
\end{enumerate}
The procedure provides the mode's energy $\omega_Q$ and the associated mass parameter $M_Q$.

It is worth mentioning that there are other methods to extract collective mass parameters that are typically defined in a vacuum~\cite{Li2022}.  
One example is the adiabatic approximation to the time-dependent Hartree-Fock-Bogoliubov approach (ATDHFB), commonly used to analyze large amplitude collective motion~\cite{ring-schuck} and widely used in nuclear systems~\cite{villars,Baranger1978,goeke}. The fundamental idea behind this method is that the collective motion of the whole system is slow compared to the independent movement of each of the nucleons that the system consists of. Based on that, one can obtain collective mass, as it has been done in~\cite{baran2011quadrupole,PhysRevC.109.L051301,PhysRevC.105.034603}. Additionally, this approach has various modifications, among which one should point out the standard method called cranking approximation~\cite{ring-schuck,Girod1979}. Introduced in~\cite{inglis} and~\cite{thouless-valatin}, the cranking method has become the standard method commonly used in collective motion examination. All these approaches extract the mass parameter from quasiparticle wave function, corresponding to static solutions of time-independent HFB equations ($i\hbar\partial / \partial 
t\rightarrow E_{\eta,q}$ in Eq.~(\ref{method:bdg})). Despite the fact that they can also be applied in the analysis of large 3D systems, as it has been shown, e.g., in~\cite{umar} and \cite{goeke-2}, such methods encounter difficulties in cases of more complex problems, like nuclei immersed in the neutron matter. 
These challenges primarily stem from the significant computational cost associated with conventional methods when applied to systems involving a large number of nucleons, typically far exceeding those encountered in isolated nuclei, which usually contain on the order of a hundred nucleons. Moreover, in such systems, it is not feasible to cleanly separate the quasiparticle wave functions into subsets corresponding to the nucleus and the surrounding medium. Instead, all nucleons must be treated on equal footing, as they are strongly coupled through nuclear interactions. The method proposed in this work is designed to address these limitations. It circumvents the need for explicit linearization or construction of collective subspaces by extracting dynamical information directly from the time evolution of suitably chosen one-body observables, which are functionals of the density. While this still requires solving the full time-dependent equations of motion, this step is no longer a practical bottleneck, thanks to the availability of efficient and well-tested numerical implementations of time-dependent Hartree-Fock~\cite{maruhn2014sky3d,Abhishek2024} and time-dependent Hartree-Fock-Bogoliubov~\cite{Jin2021,pecak2024WBSkMeff} solvers tailored for large-scale nuclear systems. 
Therefore, the method is universal and can be applied to the region of the inner crust of a neutron star to extract the effective mass parameters associated with various collective degrees of motion. 

Equations of motion that arise from modern energy functionals for nuclear systems typically have a structure formally equivalent to time-dependent Hartree-Fock-Bogoliubov (HFB) equations (position and time dependence have been omitted for brevity)
\begin{align} 
    i\hbar \frac{\partial}{\partial t} 
    \begin{pmatrix}
    u_{\eta,q\uparrow}  \\
    u_{\eta,q\downarrow} \\
    v_{\eta,q\uparrow} \\
    v_{\eta,q\downarrow}
    \end{pmatrix}
    = 
    \begin{pmatrix}
    h_{q\uparrow \uparrow}  & h_{q\uparrow \downarrow} & 0 & \Delta_q \\
    h_{q\downarrow \uparrow} & h_{q\downarrow \downarrow} & -\Delta_q & 0 \\
    0 & -\Delta_q^* &  -h^*_{q\uparrow \uparrow}  & -h^*_{q\uparrow \downarrow} \\
    \Delta_q^* & 0 & -h^*_{q\downarrow \uparrow} & -h^*_{q\downarrow \downarrow} 
    \end{pmatrix}
    \begin{pmatrix}
    u_{\eta,q\uparrow} \\
    u_{\eta,q\downarrow} \\
    v_{\eta,q\uparrow} \\
    v_{\eta,q\downarrow}
    \end{pmatrix}.
    \label{method:bdg}
\end{align}
Index $q\in\{n,p\}$ stands for isopin labeling either neutron or proton. Arrows indicate the spin, and $\eta$ is the quantum number labeling the wave functions. The diagonal part of the Hamiltonian matrix contains mean-field contribution, defined via functional derivative $U_q(\bm{r},t)={\delta E}/{\delta \rho_q}$. Thus, the equations of motion arising from the modified functional~(\ref{eq:Eprimet}) account for simple modification ($\sigma \in \{\uparrow,\downarrow\}$)
\begin{equation}
    h_{q\sigma\sigma}\rightarrow h_{q\sigma\sigma}+k\left(Q(t)-Q_0\right)\frac{\delta Q}{\delta \rho_q}.
    \label{eq:UQ}
\end{equation}

The method described above is generic and does not depend on specific functional details. In this study, we use the Brussels-Montreal class of Skyrme-type functionals (BSk in short), which are particularly well suited for studies in the context of neutron stars. These functionals are developed using not only experimental data biased on finite-size nuclei but also many-body results such as pairing gaps \cite{cao2006screening} that account for infinite nuclear matter. In this way, the bulk properties are incorporated into the model. In this research, the BSk31 parametrization with microscopically deduced pairing has been applied \cite{goriely2009skyrme, goriely2016further}, which incorporates realistic pairing gaps and includes self-energy effects. The Brussels-Montreal-Skyrme functional has a generic form:
\begin{align} 
    \Ebsk(t) = \int \Epsilonbsk(\bm{r},t)\,d\bm{r}
\label{method:Ebsk}
\end{align}
with 
\begin{align} 
    \Epsilonbsk
    [\rho_q, \nabla \rho_q, \tau_q, \bm{j}_q, \nu_q]
    = \nonumber \\ 
    \sum_{q}\frac{\hbar^2}{2m_q}\tau_q + \mathcal{E}_{\rho} + \mathcal{E}_{\Delta \rho} +  \mathcal{E}_\tau + \mathcal{E}_\pi + \mathcal{E}_C.
\label{method:bsk}
\end{align}
In this functional, one can distinguish the kinetic term ($m_q$ being the bare mass of a given nucleon), $\mathcal{E}_{\rho}$ describes the properties of uniform nuclear matter at certain density, $\mathcal{E}_{\Delta \rho}$ accounts for corrections coming from nonuniformities (so-called gradient terms), $\mathcal{E}_\tau$ contributes to the effective mass of nucleon, $\mathcal{E}_\pi$ is connected with pairing correlations and $\mathcal{E}_C$ refers to Coulomb interaction. The explicit form of these terms is given in Ref.~\cite{pecak2024WBSkMeff}. In the context of this work, it is sufficient to mention that they depend on densities, which are defined through Bogoliubov amplitudes
\begin{align}\label{eq:densities-rho}
 \rho_q(\bm{r},t)  &= \sum_{\eta\sigma} \left|v_{\eta,q\sigma}(\bm{r},t)\right|^2        \\
 \tau_q(\bm{r},t)  &= \sum_{\eta\sigma}  \left|\bm{\nabla} v_{\eta,q\sigma}(\bm{r},t)\right|^2 ,  \label{eq:densities-tau}   \\
 \nu_q(\bm{r},t)   &= 2\sum_{\eta} \left[ u_{\eta,q\uparrow}(\bm{r},t) v^*_{\eta,q\downarrow}(\bm{r},t) - u_{\eta,q\downarrow}(\bm{r},t) v^*_{\eta,q\uparrow}(\bm{r},t)
                 \right],   \label{eq:densities-nu}    \\
 \bm{j}_q(\bm{r},t) &= \sum_{\eta\sigma} \mathrm{Im} \left[ v_{\eta,q\sigma}(\bm{r},t) \bm{\nabla}  v^*_{\eta,q\sigma}(\bm{r},t) \right ] .   
 \label{eq:densities-j}             
\end{align}
The densities in the above equations are the particle number density ($\rho_q$), kinetic density ($\tau_q$), anomalous density ($\nu_q$), and current density ($\bm{j}_q$). 
The summation is taken over the quasiparticle states whose energy $E_{\eta,q}$ is smaller than the cutoff energy $0<E_{\eta,q}<E_{\textrm{cut}}$. All elements in the HFB matrix are expressed via suitable functional derivatives of the energy density functional with respect to these densities. For example, the diagonal terms are 
\begin{equation}
    h_{q\sigma\sigma}=- \bm{\nabla}\frac{\delta E}{\delta\tau_q}\bm{\nabla} 
 + \frac{\delta E}{\delta\rho_q} 
 - \frac{i}{2} \left\{ \frac{\delta E}{\delta\bm{j}_q}, \bm{\nabla} \right\}, \\
\end{equation}
where $\{,\}$ is the anticommutattor. 

In this work, we will be considering oscillations of a collective variable only around its ground state value. It means that $E(t)$ from Eq.~(\ref{eq:Et}) is equivalent to $\Ebsk(t)$. However, in general, one can consider oscillations around arbitrary values of the collective variable by considering a constrained functional, like
\begin{equation}
    E = \Ebsk -\lambda_{Q}\left( Q-Q_0\right), 
    \label{eq:ELagrangeQ}
\end{equation}
where $\lambda_{Q}$ is the Lagrange multiplier assuring that the functional has a minimum at the requested point in the collective space. Next, the oscillations may be damped, which reflects the finite lifetime of various modes. To improve the quality of the mass parameter determination, we use the model of the damped harmonic oscillator when extracting the frequency of the oscillations: 
\begin{equation}\label{eq:fit}
    Q(t)=A e^{-\gamma t}\sin(\omega_{\textrm{d}} t + \varphi),
\end{equation}
where $\gamma$ is the damping coefficient, and $A$ the initial amplitude of the perturbation. Note that we take into account the correction to the frequency of damped oscillations $\omega_{\textrm{d}}$
\begin{equation}\label{eq:omega}
    \omega^2=\omega_{\textrm{d}}^2+\gamma^2.
\end{equation}
The undamped $\omega$ frequency is next used in Eq.~(\ref{eq:omegak}) for obtaining the mass parameter $M_Q$. 

\section{Results}
\label{sec:results}
We apply the method to a specific nuclear system: the inner crust of a neutron star. In this region, the crystalline lattice of bound nuclei is permeated by superfluid neutrons. This configuration naturally leads to a deeper mean-field proton potential and a higher concentration of neutrons near protons. Since these exotic neutron-rich nuclei lack a well-defined radius and have diffuse surfaces, defining neutron-based observables presents methodological challenges. However, we assume that the overall behavior of the nucleus can be characterized by tracing the proton component, which serves as the fundamental basis for measuring collective observables.

\subsection{Center of mass oscillations}
The inner crust of a neutron star is an example of a many-body polaron system~\cite{Baroni2024}, where a bound system (nucleus) is immersed in and interacts with a bath of superfluid neutrons. 
As extensively discussed in~\cite{pecak2024WBSkMeff}, extracting the effective mass of such a self-bound object is a very complex problem. On the other hand, understanding the effective mass of nuclei in the inner crust is crucial for estimating, for example, the  crystalline lattice phonon spectrum or thermal and electric conductivities~\cite{PhysRevE.64.057402,MNRASGnedin}.

The effective mass of the impurity describes its ability to move in the medium. In principle, it can be spatially dependent, and in an anisotropic medium, it can form a tensor. Since our setup is spatially uniform, and we neglect the potential effect of the bands from the crystalline lattice~\cite{chamel2005band}, the effective mass is a scalar. We will consider oscillations of the center of mass along $z$ direction
\begin{equation}
    \Zcm(t) = \frac{1}{N_p} \int z \ \rho_p(\bm{r},t) \ d\bm{r},
    \label{method:z}
\end{equation}
where $N_p=\int\rho_p(\bm{r}) d\bm{r}$ is the number of protons in the cluster, $z$ is the coordinate of the impurity along $z$ axis, $\rho_p(\bm{r})$ is the density of protons. 
By considering the position of protons only, we avoid problems with defining which neutrons belong to the impurity and which to the background. The associated contribution to the mean-field, entering the single particle Hamiltonian, according to Eq.~\eqref{eq:UQ}, is 
\begin{equation}
\begin{split}
    \Up = \frac{\delta E}{\delta \rho_p} = \UpBsk + k\left(\Zcm(t)-Z_0\right)  \frac{z}{N_p}.
    \label{eq:UpCM}
\end{split}
\end{equation}
One can recognize that the new term in the mean field corresponds to a force (computed as the gradient of the potential) which is constant in space, but the strength is time-dependent. It becomes evident that such a force can move a cluster of protons in space, but cannot deform their density distribution. 

In order to examine the system described above, we did systematic studies for average nuclear densities $\bar\rho$ ranging from $0.002 \fm^{-3}$ to $0.051 \fm^{-3}$, where the average density is defined as $\bar\rho=(N_n+N_p)/V_{WS}$, ie., number of nucleons (protons and neutrons) per Wigner-Seitz cell volume~\cite{pearson2018}. We have generated initial states, as shown in Fig.~\ref{fig:sketch}, by means of solving the static variant of problem Eq.~(\ref{method:bdg}). The problem was solved on a spatial mesh of size $N_x\times N_y \times N_z = 32 \times 32 \times 64 $ with lattice spacing $\Delta x=\Delta y=\Delta z=1.25$ fm. The associated cutoff energy is fixed through the value of resolution to $E_{\mathrm{cut}} = \hbar^2 \pi^2/(2 m_n \Delta x^2) \approx 130 $ MeV.
In the computation, we have used \verb|W-BSK Toolkit|, described extensively in Ref.~\cite{pecak2024WBSkMeff}, and available as an open-source package via webpage~\cite{WBSKToolkit}. Next, we have performed time-dependent simulations with modified mean-field potential Eq.~(\ref{eq:UpCM}) with the initial displacement $Z_0=1$fm. 
\begin{figure}[t]
 \includegraphics[width=\columnwidth]{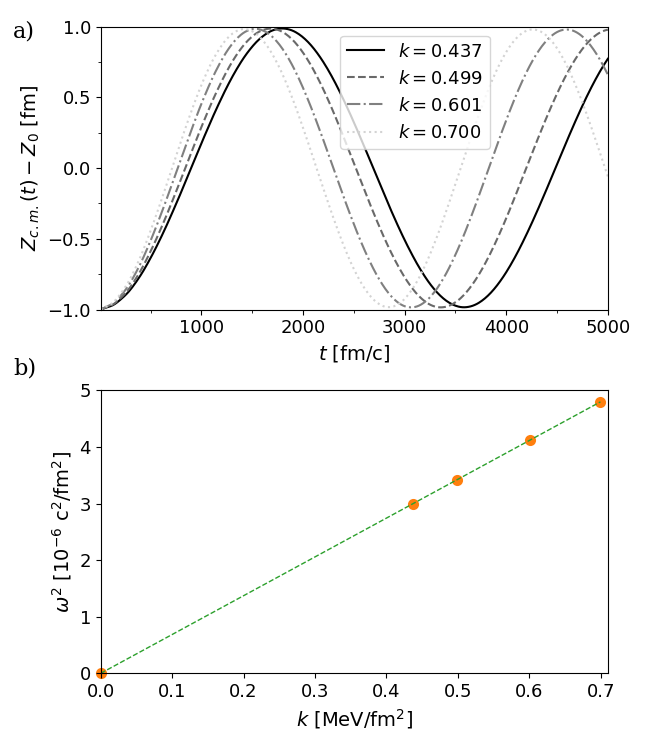}
 \caption{a)~Position of the center of mass of protons as a function of time for different values of the $k$ parameter. The considered system has average nuclear density $\bar\rho = 0.015 \fm^{-3}$. b)~The extracted frequency of oscillations as a function of the $k$ parameter. The dashed line is the fit of Eq.~(\ref{eq:omegak}), $\omega^2 = 7\cdot 10^{-6} k$.
 \label{fig:2}
 }
\end{figure}
We note that the simulations are performed with periodic boundary conditions, whereas the potential defined in Eq.~(\ref{eq:UpCM}) does not strictly satisfy this constraint, i.e., $\Up(x,y,z) \neq \Up(x,y,z + N_z \Delta z)$. However, this does not pose an issue in our case, as $\Up$ is coupled solely to the proton density, which is spatially localized and effectively vanishes at the boundaries of the simulation box.

It should be noted that the spin–orbit contribution to the nuclear interaction has been neglected in the present work. While this term is known to play a crucial role in the structure and dynamics of finite nuclei, its impact in the neutron star crust is markedly reduced. Previous studies (e.g., \cite{pearson2018}) have shown that the spin–orbit interaction has little influence on the results in this regime. Moreover, as discussed in \cite{pecak2024WBSkMeff}, spatial density fluctuations in the crust are expected to be small, and in the limiting case of uniform matter, the spin–orbit term vanishes entirely. Neglecting this term, therefore, represents a reasonable approximation in the present context and, importantly, reduces the computational cost, enabling simulations of sufficiently large volumes relevant for neutron star crust studies. We emphasize that the proposed method is fully general and can be applied regardless of whether the spin–orbit term is included in the underlying energy density functional.

We applied the fitting procedure to extract the effective parameters of the impurity. 
Fig.~\ref{fig:2} shows the center of mass oscillations for different values of control parameter $k$, for selected nuclear density $\bar\rho = 0.015 \fm^{-3}$. As expected, the extracted frequencies of oscillations follow Eq.~\eqref{eq:omegak}. One can clearly see that the intercept in this case is $\omega_{\Zcm}=0$. This feature of the center of mass collective degree of freedom is also expected. Since the Hamiltonian is spatially symmetric, i.e., it is invariant under the operation $\bm{r}\rightarrow\bm{r}+\bm{r}_0$, shifting the nucleus does not change the energy of the system. Nevertheless, the translational symmetry is broken: the nuclear cluster can form at any place, which is set randomly. The $\omega_{\Zcm}=0$ mode is then analogous to a zero-energy Goldstone mode associated with the breaking of translational symmetry. This mode does not represent oscillation but rather a shift with a constant velocity that moves the ground state across all degenerate quantum states that are also solutions of the Hamiltonian.

\begin{figure}
 \includegraphics[width=\columnwidth]{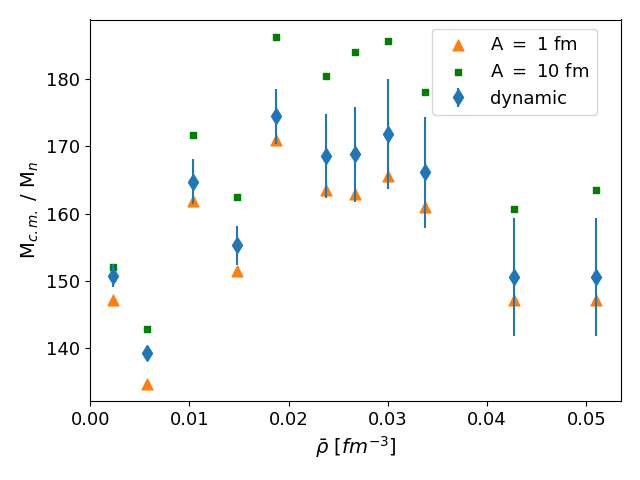}
 \caption{The effective mass $M_\cm$ (in units of neutron mass $m_n$) as a function of mean nuclear density $\bar\rho$ extracted for small amplitude $A=1$ fm (orangle triangles) and large amplitude $A=1$ fm (green squares) oscillations. We compare the results of the proposed method (orange triangles) with the effective mass obtained recently using a different dynamical scheme (blue diamonds)~\cite{pecak2024WBSkMeff}.
 }
 \label{fig:3}
\end{figure}
The extraction of the effective mass of the nucleus as a function of the inner crust mean nuclear density $\bar\rho$ is summarized in Fig.~\ref{fig:3}. The results are benchmarked against independent calculations performed within the same numerical framework~\cite{pecak2024WBSkMeff} but based on a different scheme. This benchmark scheme is based on the measurement of acceleration once a constant external force is applied to the nucleus. It is specific to the center-of-mass collective variable and cannot be extended to other collective variables, a limitation that our current approach overcomes. The results are consistent within numerical uncertainties, which remain below 5\%.

\begin{figure}[t]
 \includegraphics[width=\columnwidth]{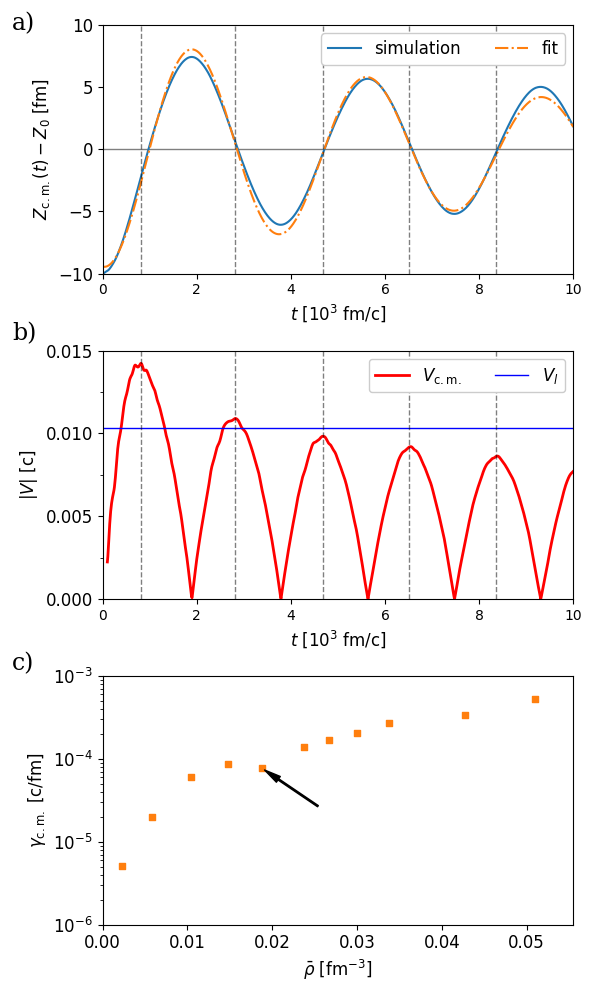}
 \caption{a) The damped oscillations of the center of mass of protons for selected density $\bar\rho = 0.019 \fm^{-3}$ (we show that point in panel c), and b)~associated absolute value of velocity as a function of time. The velocity is compared to Landau's velocity. In panel c), we show the extracted damping coefficient $\gammaCM$ as a function of average nuclear density. 
 \label{fig:4}
 }
\end{figure}
The extraction of the mass parameter is valid in the limit of small-amplitude oscillations. In this regime, no damping is observed (see Fig.~\ref{fig:2}a), as the nucleus moves through the superfluid without resistance. This behavior reflects a fundamental property of superfluids: as long as the velocity remains below a critical threshold $\Vlandau$, the system does not exert a drag force on the moving object. To explore dissipative effects, the numerical experiment can be readily adjusted to achieve higher maximum velocities by increasing the oscillation amplitude $A$. The resulting maximum velocity is proportional to  $\max\left[V_{\textrm{CM}}\right]=A\omega_d$, where $V_{\textrm{CM}}(t) = \dot{Z}_{\textrm{CM}}$. 
We have repeated the simulations with an increased value of amplitude $A=10$ fm, which is of the same order as the radius of the impurity. For the new setup, the maximum velocity exceeds this so-called Landau velocity 
\begin{equation}\label{eq:vl}
\Vlandau = \frac{\Delta_n}{\hbar {\kF}_n},    
\end{equation}
where $\Delta_n$ and ${\kF}_n$ are the bulk values of the pairing field and the Fermi wave-vector of neutrons. Note that this quantity is defined for a uniform system, and it is not obvious how well it works for such a non-uniform system as we study. For increased amplitude, we observed damped oscillations as shown in Fig.~\ref{fig:4}a). From fitting to the Eq.~\eqref{eq:fit} we can accurately reproduce the behaviour of damped oscillations (orange solid line) and extract the damping parameter $\gammaCM$. 
We investigate the absolute value of the velocity of the impurity in Fig.~\ref{fig:4}b). One can clearly connect the damping with exceeding the Landau velocity by the impurity. The damped oscillator model starts to underestimate the oscillations' amplitude once the maximum velocity drops significantly below the Landau velocity.
A similar mechanism was observed in recent work~\cite{pecak2024WBSkMeff} where the breaking of Cooper pairs was one of the mechanisms of energy dissipation.
The last panel of Fig.~\ref{fig:4}c) shows the damping parameter $\gammaCM$ as a function of average nuclear density in the inner crust. This is a monotonically increasing function of the density. In the limit of low density $\bar{\rho}\rightarrow 0$, it corresponds to a nucleus oscillating in the vacuum. Obviously, in this regime, there are no degrees of freedom where the energy can be dissipated, thus no damping is present. Once the medium gets denser, we see that the damping coefficient increases as well. 
This effect comes from the investigation of Eq.~\eqref{eq:vl} where the pairing field $\Delta_n$ vanishes in high densities \cite{cao2006screening}, while the wave vector in the denominator $\kF$ grows. 
Note that in this work, we consider only the ${}^1S_0$ pairing channel, and neglect other channels for the pairing, such as higher angular momentum channels ${}^3P_1$, which is believed to dominate in higher densities.

The case of oscillations with amplitude $A = 10$ fm cannot be considered small, as the amplitude becomes comparable to, or even exceeds, the size of the protonic cluster itself, whose radius ranges from $5$ to $10$ fm depending on the density. Nevertheless, extracting the mass parameter from these data still yields reasonable results, see square markers in Fig.~\ref{fig:3}. Although the method consistently overestimates $\Mcm$ by approximately $10-15$\%, the qualitative behavior is well captured. This discrepancy can be viewed as an estimate of the upper bound on the method’s uncertainty.

\subsection{Quadrupole moment oscillations}
The center-of-mass collective variable of a spherical nucleus provides the simplest example, serving as an ideal benchmark against existing results in the literature. A slightly more complicated case involves the collective variable associated with nuclear deformation.
As the next example, we therefore consider axial quadrupole oscillations of nuclei immersed in a sea of superfluid neutrons.
Collective modes, including quadrupole modes, are gapless, meaning they allow for excitations that can contribute, for example, to the heat capacity and transport properties of the inner crust, features that are vital for modeling the thermal evolution of neutron stars~\cite{PhysRevC.84.045801,Urban2015}.
Hydrodynamic models perform coarse-graining of physical quantities and are valid in the limit of small quasimomenta, making our microscopic approach complementary to such frameworks.
Apart from the hydrodynamic approach, quadrupole cluster vibrations in the inner crust of neutron stars have also been studied using the quasiparticle random phase approximation (QRPA), with the main emphasis placed on extracting the mode's energy~\cite{PhysRevC.71.042801,PhysRevC.99.045801}.

We define the collective variable as follows
\begin{equation}
    Q_{20}(t) = \int (3z^2-r^2) \rho_p(\bm{r},t) \ d\bm{r}.
    \label{method:beta}
\end{equation}
It vanishes for the spherical shape, is negative for oblate, and positive for prolate shapes. Division into nucleus and superfluid neutrons would be artificial. Therefore, to avoid any issues with classifying those neutrons into one belonging to the superfluid medium and those in the cluster, we define again the collective variable using only the proton density. We consider quadrupole oscillations around the spherical shape, initializing the dynamics with a deformed nucleus. Specifically, we prepare initial states with a small nonzero initial value of $Q_{20}(t=0)\approx 100\,\fm^{2}$, while the typical deformation of a heavy nucleus is of order $10^3-10^4\,\fm^{2}$. Technically, we achieve this by solving the static problem with a Lagrange constraint as in Eq.~\eqref{eq:ELagrangeQ}, within a box of size $N_x \times N_y \times N_z = 32 \times 32 \times 32$ and the lattice spacing $\Delta x = \Delta y = \Delta z = 1.25$ fm. Once the initial state is obtained, we perform the time-dependent simulations with a properly adjusted single-particle mean-field potential for protons
\begin{equation}
\begin{split}\label{method:beta-u}
    \Up = \UpBsk + k Q_{20}(t)(3z^2-r^2).
\end{split}
\end{equation}
Applying such potential results in oscillations between oblate and prolate shapes consecutively. As in the case of center-of-mass oscillations, the lack of strict periodicity in $\Up$, as required by our code, does not pose a problem, since the potential couples only to the proton density, which vanishes at the boundaries of the simulation domain.

\begin{figure}[t]
 \includegraphics[width=\columnwidth]{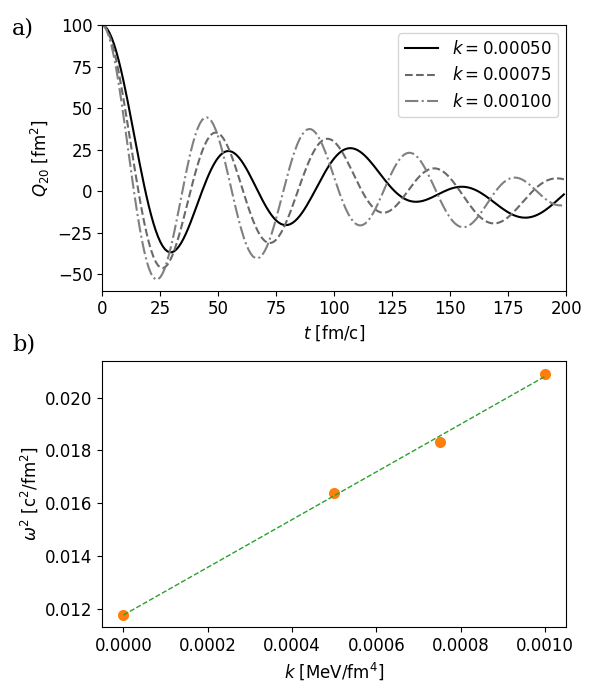}
 \caption{a) Oscillations of the quadrupole moment $Q_{20}$ as a function of time for different values of the $k$ parameter. The considered system has average nuclear density $\bar\rho = 0.015 \fm^{-3}$. b) The extracted frequency of oscillations as a function of the $k$ parameter. The dashed line is the fit of Eq.~(\ref{eq:omegak}), $\omega^2(k)=(0.1023)^2 + \frac{k}{0.0973}$. }
 \label{qm-example}
\end{figure}
Representative results for the selected density $\bar\rho = 0.015 \,\fm^{-3}$ are shown in Fig.~\ref{qm-example} for various values of the control parameter $k$. In all cases, we observe that the quadrupole mode is damped, even in the limit of small-amplitude oscillations, indicating its finite lifetime. Nevertheless, the time evolution of $Q_{20}(t)$ is well described by a damped harmonic oscillator model, allowing the extraction of the oscillation frequency. The dependence of the frequency on the parameter $k$ follows the expected behavior given by Eq~\eqref{eq:omegak}. It is worth noting that, unlike the center-of-mass oscillations, the quadrupole mode is a finite-energy excitation, with $\hbar \omega_{Q_{20}}>0$.

\begin{figure}[t]
 \includegraphics[width=\columnwidth]{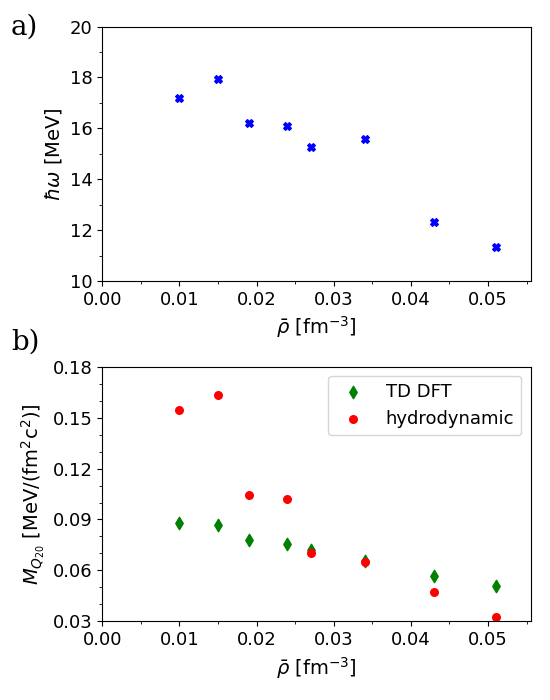}
 \caption{a) The energy of the quadrupole mode excitations as a function of the inner crust density $\bar\rho$. 
 b) The mass parameter $M_{Q_{20}}$ calculated with TDDFT (green diamonds) and the hydrodynamic approach (red circles). 
 \label{fig:6}
 }
\end{figure}
We extract the oscillation frequencies systematically for various depths of the inner crust. In Fig.~\ref{fig:6} we present results for the mode's energy $\hbar\omega_{Q_{20}}$ and the associated mass parameter $M_{Q_{20}}$ as a function of the mean density $\bar\rho$ of the system. The $Q_{20}$ excitation energy decreases with the density. Furthermore, we see that the mass parameter $M_{Q_{20}}$ is also decreasing. This quantity shows how easy it is to deform a nucleus. We observe that the presence of background neutrons diffuses the border of the impurity and allows for deformation at a lower energy cost. It is understandable, as it is known that by increasing density, the protons cease forming spherical clusters in neutron matter but instead evolve into exotic shapes known under the common name \textit{nuclear pasta}~\cite{RevModPhys.89.041002}. The spherical clusters predominantly transform into rod-like structures, which inevitably requires increasing the quadrupole moment to large values at a low energy cost. It is also interesting to note that the simple linear extrapolation of the mass parameter and the mode's energy points at the density $\approx 0.09 \fm^{-3}$ as the value around which it vanishes. This density is close to the expected density at which the transition between the crust and the core happens~\cite{haensel2007}.

The mass parameter can be estimated with the help of  the hydrodynamic approach (see Appendix~\ref{appendix} for derivation of the formula):
\begin{equation}
    M_{Q_{20}} = \frac{16\pi m_n^3}{5M_\textrm{CM}^2} \rho_\textrm{out} (\kappa - 1)^2 \frac{R_0}{2\kappa+3},
    \label{eq:hydro-MQ}
\end{equation}
where $\kappa=\rho_\textrm{in}/\rho_\textrm{out}$, with $\rho_\textrm{out}$ being density of neutrons far from the impurity and $\rho_\textrm{in}$ being density of protons and neutrons inside the impurity. 
We compare this prediction with our results in Fig.~\ref{fig:6}b). We find that for higher densities, the results given by the hydrodynamic method stand in good agreement with the results that we obtained. On the other hand, for densities below $0.03 \fm^{-3}$, the values obtained from hydrodynamics overestimate our predictions, in the low-density limit by a factor of about two. The discrepancies arise, e.g., from the estimation of the radius of impurity --- in the hydrodynamical model, we assume a sharp radius of the impurity. 

\section{Conclusions}
\label{sec:conclusions}
We have introduced a new method based on time-dependent density functional theory for extracting mass parameters. This approach is broadly applicable to a wide range of physical systems and collective variables, provided that the variable of interest can be expressed in terms of the density. In principle, this poses no limitation, as the foundational theorems of density functional theory guarantee that all observables can be written as functionals of the density. In practice, however, the explicit functional dependence may not be known for some observables.

The method is well-suited to complex scenarios, such as those encountered in studies of neutron star interiors, where crystalline nuclear structures coexist with superfluid neutrons. Standard techniques often become impractical in such contexts, but the proposed approach offers a promising alternative.
As a demonstration, we applied the new method to two cases: extracting the effective mass of nuclear clusters embedded in superfluid neutron matter, and determining the mass parameter associated with quadrupole oscillations. For the former, our results are consistent with earlier findings~\cite{pecak2024WBSkMeff}. For the latter, the extracted mass parameter provides new insights into the system's behavior, characterizing the inertia of an impurity with respect to quadrupole deformations. Notably, the results reveal the increasing significance of this mode in lattice vibrations as the density increases.

The development of new methods grounded in density functional theory lays the foundation for constructing effective models of complex systems such as neutron stars. In particular, combining static and time-dependent approaches appears to be especially powerful. Some quantities, such as the effective potential $V(Q)$ in Eq.(\ref{eq:HQ}), are most efficiently extracted using static methods. Others, typically associated with dynamical behavior, are better addressed using time-dependent approaches. Here, we demonstrated the latter by extracting the mass parameter $M_{Q_{20}}$. Together, these complementary methods can provide robust input for collective models of neutron stars~\cite{PhysRevC.87.055803,PhysRevC.90.065805,PhysRevC.84.045801}, which are essential for linking microscopic properties to macroscopic observational data.

A realistic effective model of a rotating neutron star will also need to incorporate quantum vortices. These topological defects are often assumed to be massless~\cite{Barenghi2006-wa}, although several studies suggest that vortices in Fermi superfluids may in fact carry mass~\cite{Kopnin1991,Kopnin1998,Simanek1995,Volovik1998,Kopnin2002,Simula2018}. The TDDFT-based method presented here is not suited for determining the effective mass of such defects. However, a recent study has proposed an approach capable of quantifying the mass of quantum vortices using TDDFT~\cite{richaud2024}.

\section*{Acknowledgements}
This work was financially supported by the (Polish) National Science Center Grants No. 2021/40/C/ST2/00072 (DP), 2021/43/B/ST2/01191 (PM, GW). 
We acknowledge Polish high-performance computing infrastructure PLGrid for awarding this project access to the LUMI supercomputer, owned by the EuroHPC Joint Undertaking, hosted by CSC (Finland) and the LUMI consortium through PLL/2023/04/016476.
\paragraph*{Author contributions:}  Research planning and construction of the method GW, AZ; numerical calculations: AZ; results analysis: AZ, DP, GW; hydrodynamic model PM, AZ. All authors contributed to the interpretation of the results and manuscript writing. 
\paragraph*{Data availability:} The data that support the findings of this article, with detailed instructions on how to reproduce the results, are openly available~\cite {zenodo}.

\appendix{}
\section{}\label{appendix}
In this section, the derivation of the formula for the mass parameter within the hydrodynamic approach is presented.

\begin{figure}[t]
 \includegraphics[width=\columnwidth]{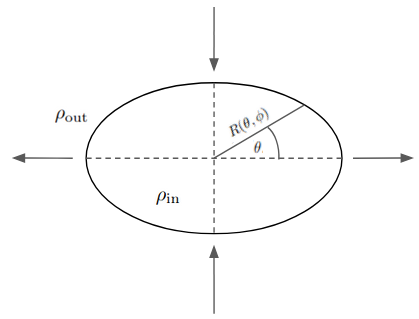}
 \caption{Schematic view of a system analyzed within the hydrodynamic approach: impurity with radius $R$ and intrinsic density $\rho_\textrm{in}$ moving in medium of density $\rho_\textrm{out}$.  
 \label{fig:appA}
 }
\end{figure}

We regard the nuclear matter as an incompressible and irrotational fluid. It implies that there exists a velocity potential $\Phi$ such that $\bm{v}=\nabla \Phi$. We consider impurity of density $\rho_\textrm{in}$ and varying with angles radius $R(\theta,\phi)$, in medium of density $\rho_\textrm{out}$, see Fig.~\ref{fig:appA}.
We define fields inside ($\Phi_\textrm{in}$) and outside ($\Phi_\textrm{out}$) of the impurity, and by utilizing spherical symmetry of the  considered system, we expand them in spherical harmonics series:
\begin{align}
    \Phi_\textrm{in} = \sum_{\lambda,\mu}\beta_{\lambda\mu} \ r^\lambda \ Y_{\lambda\mu}, \label{eq:hydro-Phi1} \\
    \Phi_\textrm{out} = \sum_{\lambda,\mu}\gamma_{\lambda\mu} \ r^{-\lambda-1} \ Y_{\lambda\mu},
    \label{eq:hydro-Phi2}
\end{align}
where $\beta_{\lambda\mu}$ and $\gamma_{\lambda\mu}$ are the expansion coefficients. This form of the expansion assures that the velocity field vanishes outside the impurity, i.e, $\Phi_\textrm{out}(r\rightarrow\infty)=0$.   
At the edge of impurity, the following boundary condition must be satisfied:
\begin{align}
\left.\Phi_{\rm in}\right|_{r=R_0} &= \left.\Phi_{\rm out}\right|_{r=R_0}, \\
\rho_{\rm in}\left( \left.\frac{\partial\Phi_{\rm in}}{\partial r}\right|_{r=R_0} - \frac{dR}{dt}\right) &=
\rho_{\rm out}\left( \left.\frac{\partial\Phi_{\rm out}}{\partial r}\right|_{r=R_0} - \frac{dR}{dt} \right).
\end{align}
For the shape parametrization (\ref{eq:Rthetapshi}) we have
\begin{equation}
    \frac{dR}{dt} = R_0 \sum_{\lambda,\mu} \dot{\alpha}_{\lambda\mu} \ Y_{\lambda\mu}(\theta,\phi),
\end{equation}
and the boundary condition reveals expressions for $\beta_{\lambda\mu}$ and $\gamma_{\lambda\mu}$ as:
\begin{align}
    \beta_{\lambda\mu} &=\gamma_{\lambda\mu}R_0^{-2\lambda-1}, \\
    \gamma_{\lambda\mu} &= R_0^{\lambda+3} \frac{\kappa-1}{\kappa \lambda + (\lambda+1)} \dot{\alpha}_{\lambda\mu}
    \label{eq:hydro-beta-gamma}
\end{align}
with $\kappa=\rho_\textrm{in}/\rho_\textrm{out}$.
Using the relation (divergence theorem)
\begin{equation}
    \int_V \left| \nabla \Phi \right|^2 d\bm{r} = \int_S \Phi \nabla \Phi \cdot d\bm{S},
\end{equation}
we can write the kinetic energy of the impurity as
\begin{align}
    T =\ & \frac{1}{2}m_n\rho_\textrm{in} \int_{r=R_0} \Phi_\textrm{in} \frac{\partial \Phi_\textrm{in}}{\partial r} R_0^2 \, d\Omega \nonumber \\
        & - \frac{1}{2}m_n\rho_\textrm{out} \int_{r=R_0} \Phi_\textrm{out} \frac{\partial \Phi_\textrm{out}}{\partial r} R_0^2 \, d\Omega.
    \label{eq:hydro-T}
\end{align}
Next, substituting the expansions~(\ref{eq:hydro-Phi1})-(\ref{eq:hydro-Phi2}), we obtain the formula :
\begin{equation}
    T = \frac{1}{2} m_n\rho_\textrm{out}(\kappa-1)^2R_0^5 \sum_{\lambda,\mu} \frac{1}{\kappa \lambda + (\lambda+1)} |\dot{\alpha}_{\lambda\mu}|^2.
    \label{eq:hydro-T-2}
\end{equation}
Comparing it with the general formula for the kinetic energy
\begin{equation}
    T = \frac{1}{2} \sum_{\lambda,\mu} M_\lambda |\dot{\alpha}_{\lambda\mu}|^2
    \label{eq:hydro-kinetic}
\end{equation}
we can identify the associated mass parameters
\begin{equation}
    M_\lambda = m_n\rho_\textrm{out}(\kappa-1)^2R_0^5\frac{1}{\kappa \lambda + (\lambda+1)}.
\end{equation}
For the quadruple mode ($\lambda=2$) it reduces to
\begin{equation}
    M_{2} = m_n\rho_\textrm{out} (\kappa - 1)^2 \frac{R_0^5}{2\kappa+3}.
    \label{eq:hydro-Malpha}
\end{equation}
The radius $R_0$ can be related to the effective mass $M_\textrm{CM}$ of the impurity 
\begin{equation}
    R_0 = \Bigg(\frac{3M_\textrm{CM}}{4\pi m_n \rho_\textrm{in}}\Bigg)^{1/3}.
    \label{eq:hydro-R}
\end{equation}
Finally, we relate the quadrupole moment to the deformation parameter
\begin{equation}
    Q_{20}=C \alpha_{20},\quad C=\sqrt{\frac{5}{16\pi}} \frac{M_\textrm{CM}}{m_n} R_0^2.
\end{equation}
Then, mass parameter for quadrupole moment oscillations $M_{Q_{20}}$ reads:
\begin{equation}
    M_{Q_{20}} = \frac{M_{2}}{C^2},
\end{equation}
which gives Eq.~(\ref{eq:hydro-MQ}) from the main text.

\section{}
Here, we provide a tabularized version of the data presented in Figs \ref{fig:3}, \ref{fig:4}c, and \ref{fig:6}. 
\begin{table*}[ht]
	\centering 
		\begin{tabular}{|c|c|c|c|c|c|c|c|} 
            \hline
            $\bar\rho$ & $\rho_{Bn}$ & $R_0$ & $\Mcm$ (A=1fm) & $\Mcm$ (A=10 fm) & $\gammaCM$ (A=10 fm) & $M_{Q_{20}}$ & $\hbar\omega_{Q_{20}}$ \\
            $[\fm^{-3}]$ & $[\fm^{-3}]$ & $[\fm]$ & $[m_n]$ & $[m_n]$& $[c\fm^{-1}]$ & $[\MeV\fm^{-2}c^{-2}]$ & $[\MeV]$ \\
			\hline\hline
			0.0023 & 0.0016 & 6.16 & 147.17 & 152.09 & 5.05e-06  & 0.104  & 21.7 \\
			0.0058 & 0.0045 & 5.96 & 134.63 & 142.78 & 2.02e-05  & 0.107  & 22.2 \\
			0.0104 & 0.0084 & 6.53 & 161.81 & 171.63 & 6.03e-05  & 0.0878 & 17.2 \\
			0.0148 & 0.0120 & 6.39 & 151.49 & 162.52 & 8.74e-05  & 0.0870 & 17.9 \\
			0.0187 &0.0152 & 6.78 & 170.91 & 186.30 & 7.77e-05  & 0.0777 & 16.2 \\
			0.0237 & 0.0193 & 6.72 & 163.46 & 180.44 & 1.38e-04  & 0.0756 & 16.1 \\
			0.0267 & 0.0217 &  6.76 & 162.98 & 184.11 & 1.70e-04  & 0.0719 & 15.3 \\
			0.0300 & 0.0244 &6.85 & 165.54 & 185.64 & 2.05e-04  & 0.0675 & 14.5 \\
			0.0338 & 0.0276 & 6.83 & 160.94 & 178.12 & 2.74e-04  & 0.0655 & 15.6 \\
			0.0428 &0.0351 & 6.73 & 147.19 & 160.73 & 3.37e-04  & 0.0567 & 12.3 \\
			0.0510 & 0.0422 & 6.84 & 147.15 & 163.48 & 5.24e-04  & 0.0508 & 11.3 \\
			\hline
		\end{tabular}
	\caption{The consecutive columns describe the layer of the inner crust of density $\bar\rho$ with bulk neutron density $\rho_{Bn}$. We provide the radius of the zirconium nucleus $R_0$, computed according to formula~(\ref{eq:hydro-R}). The effective masses $\Mcm$ are provided for runs with amplitudes of oscillations equal to 1fm and 10fm.  $M_{Q_{20}}$ indicates the mass parameter related to the quadrupole deformation. The damping coefficient for A=10fm oscillations is given by $\gammaCM$. The quadrupole mode energy $\hbar\omega_{Q_{20}}$ is depicted in that last column. The number of bound neutrons slightly depends on the amplitude. However, it can be easily  obtained by subtracting the number of protons (Z=40) from the column $\Mcm$.}
	\label{table}
\end{table*}

\bibliography{bibtexNS}
\end{document}